# A novel 3D multi-path DenseNet for improving automatic segmentation of glioblastoma on pre-operative multi-modal MR images

Jie Fu[1], Kamal Singhrao[1], X. Sharon Qi[1], Yingli Yang[1], Dan Ruan[1], and John H. Lewis[2*]

1. Department of Radiation Oncology, University of California, Los Angeles, Los Angeles, CA, USA, 90095

2. Department of Radiation Oncology, Cedars-Sinai Medical Center, Los Angeles, CA, USA, 90048

*Email: john.h.lewis@cshs.org

## Abstract

Convolutional neural networks have achieved excellent results in automatic medical image segmentation. In this study, we proposed a novel 3D multi-path DenseNet for generating the accurate glioblastoma (GBM) tumor contour from four multi-modal pre-operative MR images. We hypothesized that the multi-path architecture could achieve more accurate segmentation than a single-path architecture. 258 GBM patients were included in this study. Each patient had four MR images (T1-weighted, contrast-enhanced T1-weighted, T2-weighted, and FLAIR) and the manually segmented tumor contour. We built a 3D multi-path DenseNet that could be trained to generate the corresponding GBM tumor contour from the four MR images. A 3D single-path DenseNet was also built for comparison. Both DenseNets were based on the encoder-decoder architecture. All four images were concatenated and fed into a single encoder path in the single-path DenseNet, while each input image had its own encoder path in the multi-path DenseNet. The patient cohort was randomly split into a training set of 180 patients, a validation set of 39 patients, and a testing set of 39 patients. Model performance was evaluated using the Dice similarity coefficient (DSC), average surface distance (ASD), and 95% Hausdorff distance ($HD_{95\%}$). Wilcoxon signed-rank tests were conducted to examine the model differences. The single-path DenseNet achieved a DSC of 0.911±0.060, ASD of 1.3±0.7 mm, and $HD_{95\%}$ of 5.2±7.1 mm, while the multi-path DenseNet achieved a DSC of 0.922±0.041, ASD of 1.1±0.5 mm, and $HD_{95\%}$ of 3.9±3.3 mm. The p-values of all Wilcoxon signed-rank tests were less than 0.05. Both 3D DenseNets generated GBM tumor contours in good agreement with the manually segmented contours from multi-modal MR images. The multi-path DenseNet achieved more accurate tumor segmentation than the single-path DenseNet. Our proposed 3D multi-path DenseNet has great potential for achieving accurate GBM tumor segmentation in clinics.

**Keywords: 3D CNN; GBM tumor segmentation; Multi-path architecture**

## 1. Introduction

Gliomas are tumors arising from glial cells, normally astrocytes and oligodendrocytes. Gliomas account for approximately 26% of all brain tumors and can be classified as grades I-IV based on histological characteristics (Ostrom *et al* 2017, Louis *et al* 2016). Glioblastomas (GBM), grade IV gliomas, are the most common malignant primary brain tumors with a median overall survival of only 15 months after diagnosis (Koshy *et al* 2012, TAMIMI and JUWEID 2017). The gold standard treatment for GBM is a maximal safe resection followed by radiotherapy with or without concurrent adjuvant chemotherapy (Stupp *et al* 2009, Niyazi *et al* 2016). As intensity-modulated radiotherapy (IMRT) and volumetric-modulated arc therapy (VMAT) can deliver a high dose of radiation to the target, while providing better dose sparing of normal tissues compared to 3D conformal radiotherapy, they have been increasingly used for treating GBM.

Accurate target delineation is critical for the IMRT and VMAT treatment planning because both techniques have sharp dose gradients between the target and normal tissues. The Radiation Therapy Oncology Group (RTOG) trial recommends using multi-modal MR images, including a T2-weighted (T2w) images or a fluid-attenuated inversion recovery (FLAIR) image and a contrast-enhancing T1-weighted (CE-T1w) image, for GBM target delineation (Cabrera *et al* 2016). Manual segmentation is not only time-consuming but also sensitive to intra-observer and inter-observer variabilities. Hence, it is essential to develop automatic segmentation methods that can perform highly reproducible and accurate GBM tumor segmentation.

Recently, many convolutional neural networks (CNNs) have achieved good performance in glioma segmentation based on multi-modal MR images. An ensemble method, called EMMA (Kamnitsas *et al* 2018), earned first place in the 2017 Brain Tumor Segmentation (BraTS) challenge (Menze *et al* 2015, Bakas *et al* 2017, 2018). EMMA consisted of seven 3D CNNs including three 3D fully convolutional networks (Long *et al* 2015), two 3D U-Nets (Çiçek *et al* 2016), and two DeepMedic models (Kamnitsas *et al* 2017). Every 3D CNN in the EMMA was built based on the encoder-decoder architecture and could achieve end-to-end mapping from four multi-model MR images to tumor contour. A novel 3D CNN with autoencoder regularization earned first place in the 2018 BraTS challenge and also had the encoder-decoder architecture (Myronenko 2019). Zhang *et al* (2018) proposed a 3D DenseNet, the 3D CNN with several dense blocks, for acute ischemic stroke segmentation and showed it achieved better performance than a 3D U-Net with residual connections. The dense block was proposed by Huang *et al* (2017) to alleviate the vanishing-gradient problem, strengthen feature propagation, and encourage feature reuse. It could reduce the number of model parameters and achieved better performance on several object recognition tasks compared to the residual block (He *et al* 2016). However, all of these 3D CNNs only employed a single-path architecture for the multi-modal MR images. In other words, the same feature extraction filters were applied to the concatenation of four MR images in the single-path architecture. We hypothesized that a multi-path architecture, where each MR image has its own set of encoding filters, could achieve better segmentation performance than the single-path architecture by capturing the image-specific features.

In this study, we proposed a 3D single-path DenseNet and a 3D multi-path DenseNet for automatically generating the GBM tumor contour from four multi-modal MR images. Both DenseNets were trained, validated, and tested using a total of 258 GBM patients. Several evaluation metrics were used to compare the ground truth and autosegmented contours. The model performance of the two DenseNets was compared using Wilcoxon signed-rank tests.

## 2. Materials and methods

### 2.1. Dataset

The 2019 BraTS challenge training set, comprised of images from 259 GBM patients, was used in this study. Each patient had four pre-operative multi-modal MR images: T1w, CE-T1w, T2w, and FLAIR images. These images were acquired with different scanners and clinical protocols from multiple institutions. Preprocessing steps of co-registration and skull-stripping were applied to all MR images (Bakas *et al* 2017). The image voxel size is 1.0 x 1.0 x 1.0 mm$^3$, and the image matrix size is 240 x 240 x 155. Labels of three tumor subregions (enhancing tumor core, non-enhancing tumor core, and edema) were manually delineated by one to four raters based on the same annotation protocol. Manual delineations were approved by expert board-certified neuroradiologists to define the ground truth labels. One patient was removed because a portion of the FLAIR image was cut off, which resulted in a total number of 258 patients in this study.

**2.2. Image preprocessing**

The manual ground truth tumor contour of each patient was acquired by fusing three tumor subregion labels. The N4ITK algorithm was applied to all MR images, except the FLAIR image, to correct intensity inhomogeneity (Tustison *et al* 2010). To save computational memory, all images and contours were cropped to exclude the background margin and resampled to have an isotropic voxel size of 1.5 x 1.5 x 1.5 mm$^3$. The final matrix size of the images and contours was 100 x 128 x 105. For each MR image, voxel intensity was normalized to z-score using the mean and standard deviation of the intensities of its brain voxels. Figure 1 shows the transverse slices of four preprocessed MR images along with the ground truth contour for one example patient.

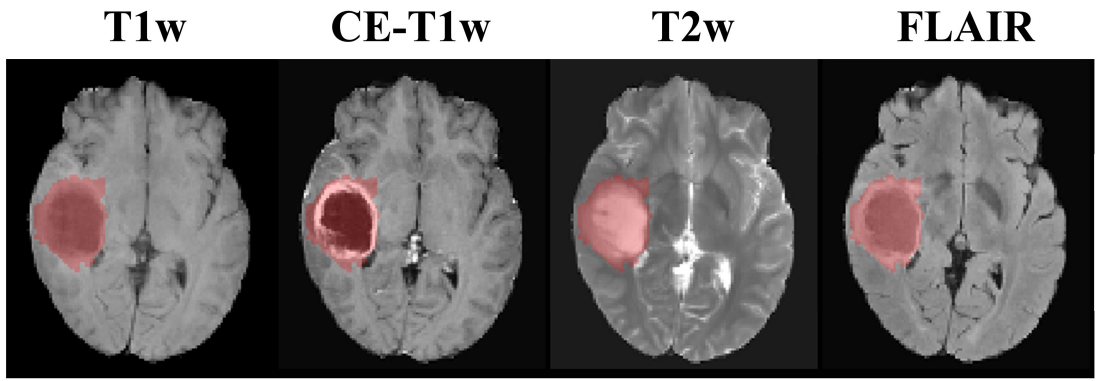

Figure 1. From left to right, transverse slices of the: preprocessed T1-weighted (T1w), contrast-enhanced T1w (CE-T1w), T2-weighted (T2w), and fluid-attenuated inversion recovery (FLAIR) MR images, along with the ground truth tumor contour for one example patient. Z-score window [-4, 4] is used for image display.

**2.3. 3D CNNs**

**2.3.1. 3D single-path DenseNet**

Figure 2 shows the architecture of a 3-layer dense block used in the proposed 3D DenseNets. Each layer in the dense block contained one convolution layer with a filter size of 1 x 1 x 1 followed by one convolutional layer with a filter size of 3 x 3 x 3. The number of output feature maps after each 1 x 1 x 1 convolutional layer is the growth rate of the dense block. Instance normalization layers were used to reduce internal covariate shifts and speed up model optimization (Ulyanov *et al* 2016).

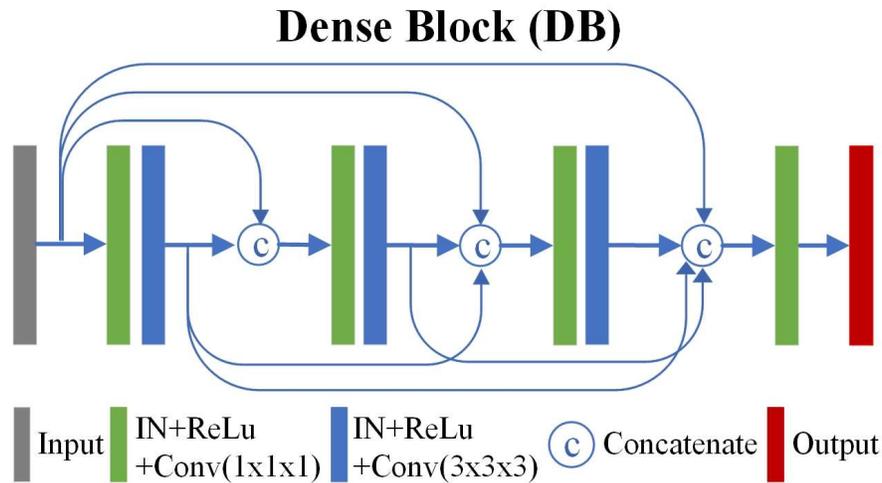

Figure 2. The architecture of the dense block. IN, instance normalization layer; ReLu, rectified linear unit layer; Conv, convolutional layers.

Figure 3 shows the architecture of the 3D single-path DenseNet for GBM tumor segmentation. It contained 5 dense blocks forming an encoder-decoder architecture similar to U-Net. The encoder path extracted features from the concatenation of four MR images, while the decoder path gradually reconstructed the contour from the extracted features. Averaging pooling layers and deconvolutional layers were used to downsample and upsample the feature maps, respectively. At the end of the model, one convolutional layer with a filter size of 1 x 1 x 1 followed by the Softmax layer was used to generate the probability maps of background and tumor labels. The model can be trained to achieve an end-to-end mapping from the concatenation of four MR images to the autosegmented tumor contour.

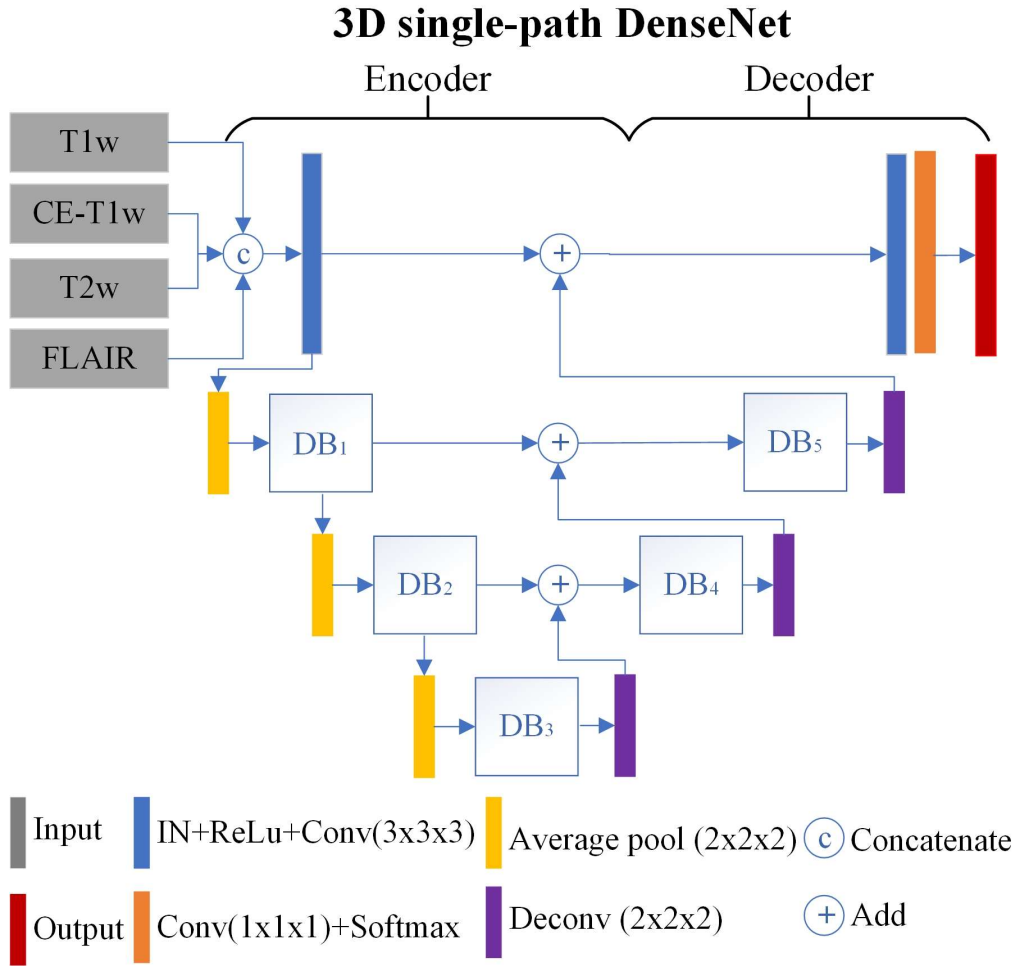

Figure 3. The architecture of the 3D single-path DenseNet. DB, dense block shown in Figure 2; IN, instance normalization layer; ReLu, rectified linear unit layer; Conv, convolutional layers; Deconv, deconvolutional layer.

### 2.3.2. 3D multi-path DenseNet

Figure 4 (A) shows the architecture of the 3D multi-path DenseNet. It also had encoder and decoder paths. In contrast to the single-path DenseNet, where four MR images were concatenated and fed into the single encoder path, each MR image has its own encoder path in the multi-path DenseNet. The encoded feature maps from four different paths were concatenated and then fused by squeeze-and-excitation blocks (SEB) as shown in Figure 4 (B). Output feature maps from the SEBs were fed into the same decoder path that was used in the single-path DenseNet. The SEB was proposed by Hu *et al* (2018) to recalibrate the channel-wise feature response by modeling the inter-channel dependence. Overall, the 3D multi-path DenseNet contains 14 dense blocks. In each

SEB, the number of output feature maps after the convolutional layer and the number of nodes in two fully connected layers were set the same as the growth rate of the dense block.

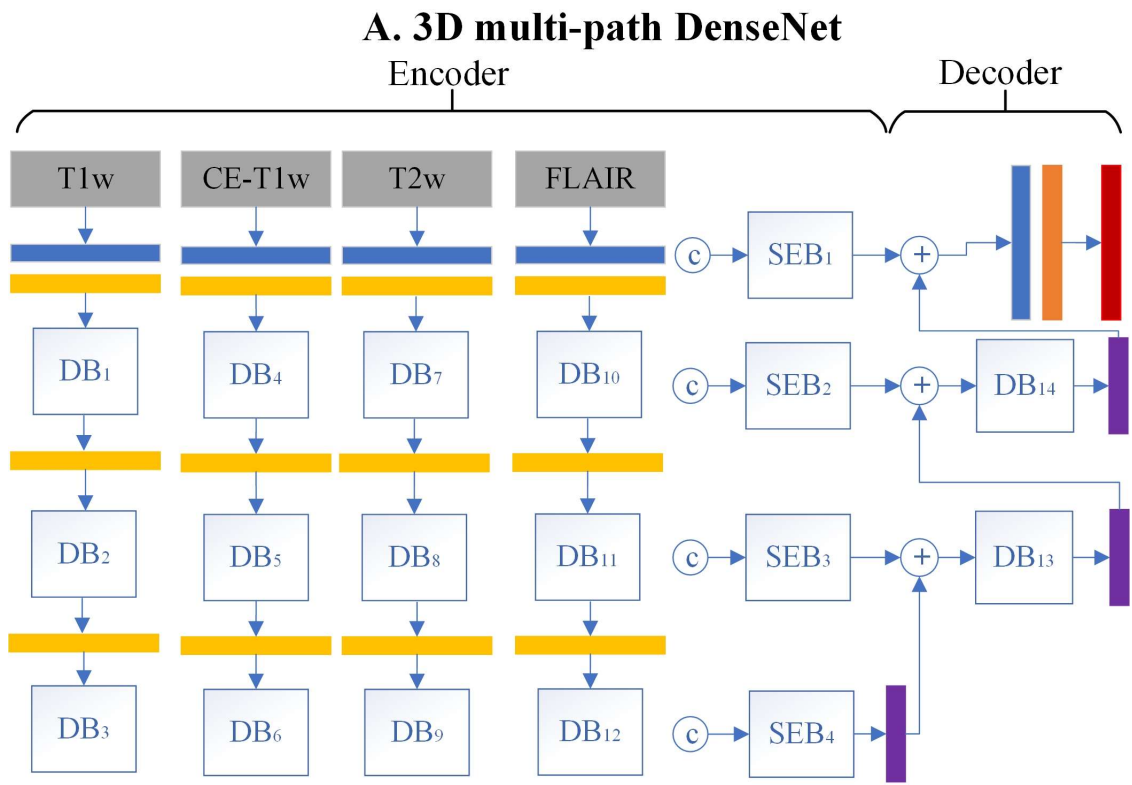

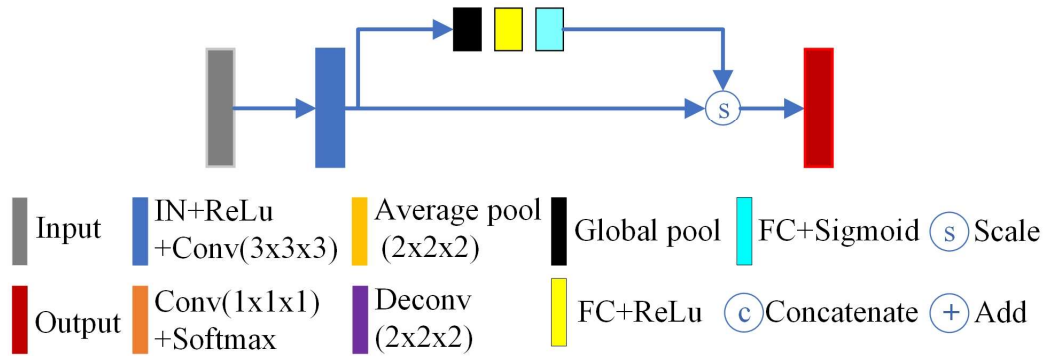

Figure 4. (A) The architecture of the 3D multi-path DenseNet. (B) The architecture of the squeeze-and-excitation blocks (SEB). DB, dense block shown in Figure 2; IN, instance normalization layer; ReLu, rectified linear unit layer; Conv, convolutional layers; Deconv, deconvolutional layer; FC, fully connected layer.

The number of trainable model parameters only depended on the growth rate of the dense block used in the model. We set the growth rate to 30 in the single-path DenseNet and 16 in the multi-path DenseNet so that both DenseNets have a similar number of trainable parameters (about 0.46 million).

## 2.4. Model training

The patient cohort was randomly split into a training set of 180 patients, a validation set of 39 patients, and a testing set of 39 patients. The Adam stochastic gradient descent method was used to minimize the loss function,

$$loss = 1 - \sum_{i=1}^{N} \frac{2P_i \times L_i}{P_i + L_i},$$

(1)

where $N$ is the number of image voxels, $P_i$ is the Softmax probability of the voxel i being a tumor voxel, and $L_i$ is the ground truth tumor label (0: background, 1: tumor) of the voxel i.

Both DenseNets were implemented using the Tensorflow package (V1.10.0, Python 3.6.9, CUDA 10.0) and ran on an 11 GB GeForce GTX 1080 Ti. A batch size of 1 was used for training. The initial learning rate and the stopping epoch number were tuned using the validation set. For both DenseNets, the optimal initial learning rate and epoch number are $5 \times 10^{-4}$ and 90, respectively.

## 2.5. Model evaluation

Trained models were applied to 39 testing patients to generate their autosegmented tumor contours. Model performance was evaluated using three metrics: Dice similarity coefficient (DSC), average surface distance (ASD), and 95% Hausdorff distance (HD$_{95\%}$). These metrics are represented by the following equations:

$$DSC = \frac{2 \left(V_{GT} \cap V_{Auto}\right)}{V_{GT} \cup V_{Auto}},$$

(2)

$$ASD = \frac{1}{2}\left(\overline{\min_{x \in S_{GT}} d(x, S_{Auto})} + \overline{\min_{x \in S_{Auto}} d(x, S_{GT})}\right),$$

(3)

$$HD_{95\%} = \frac{1}{2}\left(K_{95}\left(\min_{x \in S_{GT}} d(x, S_{Auto})\right) + K_{95}\left(\min_{x \in S_{Auto}} d(x, S_{GT})\right)\right),$$

(4)

where $V_{GT}$ and $V_{Auto}$ refer to the volumes of the ground truth and autosegmented tumor contours, respectively; $S_{GT}$ and $S_{Auto}$ refer to the surfaces of the ground truth and autosegmented tumor contours, respectively; $\min_{x \in S_{GT}} d(x, S_{Auto})$ denotes the distance of the voxel x, on the tumor surface $S_{GT}$, to its closet voxel on the surface $S_{Auto}$; $K_{95}$ refers to the 95th percentile of all distances.

Wilcoxon signed-rank tests were conducted to compare the performance of the 3D single-path and multi-path DenseNets.

## 3. Results

Figure 5 shows the ground truth and autosegmented tumor contours for the three example patients. Autosegmented tumor contours generated by both DenseNets were similar to the corresponding ground truth tumor contour based on visual inspection. In Figure 5, white arrows point to the regions where there are larger differences between the ground truth and Auto$_{single-path}$ contours compared to those between ground truth and Auto$_{multi-path}$ contours.

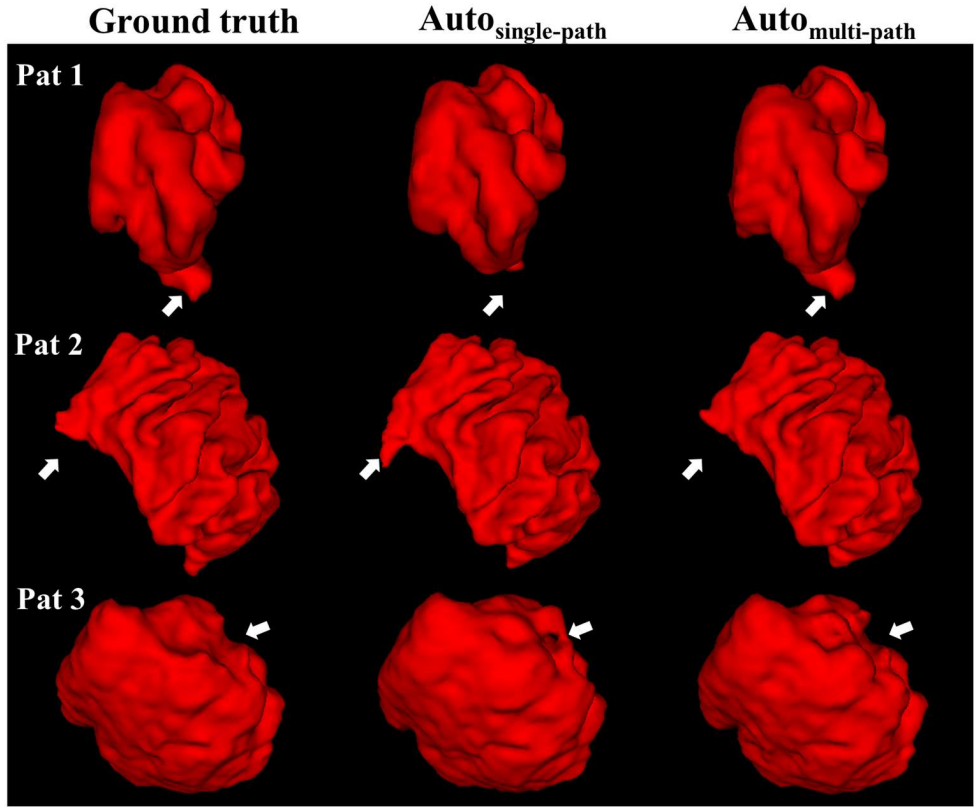

Figure 5. Ground truth tumor contours (left column) and the autosegmented tumor contours generated by the single-path DenseNet (middle column) and multi-path DenseNet (right columns) for the three example patients.

Table 1 summarizes the statistics of the evaluation metrics for the single-path and multi-path DenseNets. The multi-path DenseNet achieved a larger mean DSC of 0.922, a smaller mean ASD of 1.1 mm, and a smaller $HD_{95\%}$ of 3.9 mm compared to the single-path DenseNet. The $p$-values of all Wilcoxon signed-rank tests were less than 0.05.

| Metric | Single-path DenseNet | Multi-path DenseNet | $p$-value |
|---|---|---|---|
| DSC | 0.911±0.060 | 0.922±0.041 | <0.001 |
| ASD [mm] | 1.3±0.7 | 1.1±0.5 | 0.002 |
| $HD_{95\%}$ [mm] | 5.2±7.1 | 3.9±3.3 | 0.046 |

Table 1. Statistics of DSC, ASD, and $HD_{95\%}$ between the ground truth contours and the autosegmented contours generated by the single-path DenseNet or multi-path DenseNet. Results were averaged across 39 testing patients and shown in (mean ± SD) format. The $p$-values of the Wilcoxon signed-rank tests are shown.

Figure 6 shows box and whisker plots of the three evaluation metrics. The multi-path DenseNet generated more robust GBM tumor contours compared to the single-path DenseNet in terms of smaller box ranges (max-min) of all evaluation metrics.

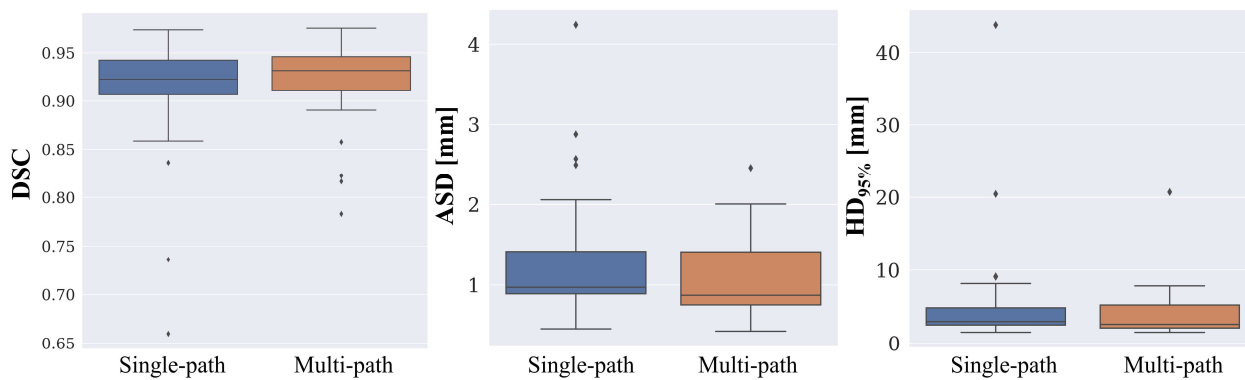

Figure 6. Box and whisker plots of DSC, ASD, and HD$_{95\%}$ for the single-path and multi-path DenseNets. The maximum (top line), 75$^{th}$ percentile (top of the box), median (central line), 25$^{th}$ percentile (bottom of the box), and minimum (bottom line) are shown. Outliers are drawn as diamond signs.

## 4. Discussion

In this study, we proposed a 3D single-path DenseNet and a 3D multi-path DenseNet for generating the GBM tumor contour from four MR images. Both DenseNets were trained, validated, and tested using 180, 39, and 39 GBM patients, respectively. Autosegmneted contours generated by both DenseNets were compared with the ground truth contours using DSC, ASD, and HD$_{95\%}$.

The multi-path architecture achieved better performance in GBM tumor segmentation than the corresponding single-path architecture. The multi-path DenseNet achieved a larger mean DSC, a smaller mean ASD, and a smaller mean HD$_{95\%}$ compared to the single-path DenseNet. Results of Wilcoxon signed-rank tests indicated significant differences in all three metrics. The autosegmented contours generated by the multi-path DenseNet were generally qualitatively more similar to the ground truth contours than those generated by the single-path DenseNet, as is illustrated by the examples in Figure 5. Figure 6 showed that the multi-path DenseNet achieved more robust segmentation compared to the single-path DenseNet.

The images and ground truth tumor contours were downsampled from the original voxel size of 1.0 x 1.0 x 1.0 mm$^3$ to the voxel size of 1.5 x 1.5 x 1.5 mm$^3$ to save computational memory. We upsampled the autosegmented contours generated by 3D DenseNets back to the original spatial resolution and compared them with original ground truth tumor contours. In this case, the single-path DenseNet achieved a mean DSC of 0.900, while the multi-path DenseNet achieved a mean DSC of 0.910. The Wilcoxon signed-rank test suggested a significant difference ($p$-value<0.001). The mean DSC results evaluated in the original spatial resolution are comparable to the mean DSC of 0.884 that was achieved by the 3D CNN with autoencoder regularization in the 2018 BraTS challenge (Myronenko 2019).

The goal of our study was to test the hypothesis that the proposed 3D multi-path DenseNet could achieve better GBM tumor segmentation than the corresponding single-path DenseNet. Our proposed multi-path technique could be integrated into other 3D CNNs that are based on the encoder-decoder architecture for improving GBM tumor segmentation. But this was not explored within the scope of this study. Also, our proposed multi-path technique may help achieve better performance in other image-transfer tasks, such as synthetic CT generation and organ-at-risk segmentation from multi-modal MR images. The image output in the proposed DenseNets can be modified to a single channel for synthetic CT generation, and multiple channels for organ-at-risk or tumor subregion segmentation. Future work will include integrating the multi-path technique into other 3D CNNs to potentially improving GBM segmentation performance, and investigating the performance of the multi-path technique in other image-transfer tasks.

## 5. Conclusion

We proposed a 3D single-path DenseNet and a 3D multi-path DenseNet for automatically generating GBM tumor contours from four multi-modal MR images. Both DenseNets generated accurate tumor contours. The single-path and multi-path DenseNets achieved DSCs of 0.911±0.060 and 0.922±0.041, respectively. Our study showed that the multi-path DenseNet generated more accurate GBM tumor contours than the single-path DenseNet.

**Acknowledgment**

This research was funded by Varian Medical Systems, Inc.

**Disclosure of Conflicts of Interest**

The authors have no relevant conflicts of interest to disclose.